\documentclass[floatfix,eqsecnum,aps,nofootinbib]{revtex4}

\usepackage{amsmath}
\usepackage{amssymb}
\usepackage{amscd}
\usepackage{mathrsfs}
\usepackage{graphicx}
\usepackage{subfigure}
\usepackage{slashed}
\usepackage{mathbbol}
\usepackage{subfigure}

\newcommand{\diag}{\mathop{\mathrm{diag}}}
\newcommand{\Tr}{\mathop{\mathrm{Tr}}}

\begin{document}

\title{Neutrino masses and sparticle spectra from stochastic superspace}
\author{Archil Kobakhidze}\email{archilk@unimelb.edu.au}
\affiliation{School of Physics, The University of Melbourne, Victoria 3010, Australia}
\author{Nadine Pesor}\email{npesor@student.unimelb.edu.au}
\affiliation{School of Physics, The University of Melbourne, Victoria 3010, Australia}
\author{Raymond R. Volkas}\email{raymondv@unimelb.edu.au}
\affiliation{School of Physics, The University of Melbourne, Victoria 3010, Australia}

\begin{abstract}
Based on the stochastic superspace mechanism for softly breaking supersymmetry, we present improved sparticle spectra computations for the minimal model and examine extensions through R-parity violation and the type-I seesaw mechanism that incorporate non-zero neutrino masses for more realistic models. Performing the calculations to two-loop accuracy, we observe a global decrease in predicted sparticle masses. However this does not affect the generic features of the minimal model outlined in our earlier work, including the characteristic light stop mass. We find stop decay channels accessible at the LHC which can be used in combination with our predicted range for the stop mixing angle to falsify the minimal model with stochastic supersymmetry. We then introduce neutrino masses and mixings consistent with experiment by including purely trilinear R-parity violating superpotential terms, resulting in a viable stochastic superspace model absent a dark matter candidate. An alternative method for generating neutrino masses, namely the type-I seesaw mechanism, is found only to be viable when the neutrino Yukawa coupling is small relative to the top Yukawa and the cut-off scale is large.
\end{abstract}

\maketitle

\section{Introduction}
A number of methods exist for softly breaking supersymmetry. These can be broadly catagorized into two groups: those that originate from a fundamental theory, where supersymmetry is spontaneously broken in a hidden sector then communicated to the visible sector by a messenger, and the purely phenomenological approach, where no explanation for the cause of supersymmetry breaking is offered.

The fundamental theoretical approaches, though providing some dynamical explanation for the mechanism of supersymmetry breaking manifesting in the observable sector, suffer from a variety problems. Gravity mediation is plagued by large sparticle induced FCNC amplitudes unless no source of flavor physics exists between the weak scale and Planck scale, while gauge mediation has CP violation issues among other significant theoretical hurdles \cite{Kolda:1997wt}. Although some work-arounds for these problems exist, there is not yet a compelling and simple fundamental model for the transmission of supersymmetry breaking to the observable sector. On the other hand, writing down the most general Lagrangian with explicit soft supersymmetry breaking terms completely fails to explain the source of supersymmetry breaking. Instead, one must painstakingly explore the vast parameter space, much of which is excluded due to flavor changing neutral current (FCNC) and CP violation problems, to find phenomenologically viable regions useful for experimental predictions.

We proposed in \cite{Kobakhidze:2008py} a mechanism for supersymmetry breaking in which the Grassmannian coordinates, $\theta$ and $\bar{\theta}$, are considered to be stochastic variables. This leads to a very constrained set of soft-breaking parameters, thus significantly improving predictability compared to the phenomenological approach. However, the underlying cause of stochasticity in the Grassmannian coordinates is not postulated, placing stochastic superspace as a sort of compromise between a fully dynamical model and the purely phenomenological approach to supersymmetry breaking.

One of the goals of this paper is to improve upon the sparticle spectra calculations of the minimal stochastic superspace model presented in \cite{Kobakhidze:2008py}. Rather than using the analytic solutions to approximations of the one-loop renormalization group equations (RGEs), we employ the use of the publicly available software, {\ttfamily SOFTSUSY3.0} \cite{Allanach:2009bv}, to account for two-loop and threshold effects. We then incorporate non-zero neutrino masses into the stochastic superspace mechanism for supersymmetry breaking in order to create more realistic models. We examine the phenomenology of models that achieve this end through R-parity violation and the type-I seesaw mechanism. 

The rest of the paper is structured as follows: we briefly review the stochastic superspace formalism in the next section and present updated sparticle spectra. We then apply the stochastic superspace formalism to R-parity violation (RPV) and type-I seesaw in order to generate non-zero neutrino masses in sections \ref{sec:Rpar} and \ref{sec:seesaw}, respectively. This is followed by concluding remarks in section \ref{sec:conclusion}.

\section{The Stochastic Superspace Formalism}
\label{sec:stochastic}
In stochastic superspace models \cite{Kobakhidze:2008py}, the Grassmannian coordinates,  $\theta$ and $\bar{\theta}$, are considered to be stochastic variables with a probability distribution,
\begin{equation}
\label{eq:probdist}
\mathcal{P}(\theta, \bar{\theta})\lvert \xi \rvert^2 = 1 + \xi^{*}(\theta \theta) + \xi (\bar{\theta} \bar{\theta}) + \lvert \xi \rvert^2 (\theta \theta) (\bar{\theta} \bar{\theta}),
\end{equation}
which is dependent on a single parameter, $\xi$, a complex number of mass dimension. This probability distribution is the most general form that satisfies Lorentz invariance. The Lagrangian in ordinary spacetime is then defined as the average of the supersymmetric Lagrangian over the probability distribution in Eq.\eqref{eq:probdist}. Averaging over the superpotential for the minimal supersymmetric standard model (MSSM),
\begin{equation}
W_{\text{SM}} = \mu H_u H_d + \hat{y}^{\text{up}} Q U^c H_u + \hat{y}^{\text{down}} Q D^c H_d + \hat{y}^{\text{lept}} L E^c H_d,
\end{equation}
gives rise to the soft-breaking terms
\begin{equation}
L_{\text{soft scalar}} = -\xi^* \mu \tilde{H}_u \tilde{H}_d - 2 \xi^* \left[\hat{y}^{\text{up}} \tilde{Q} \tilde{U}^c \tilde{H}_u + \hat{y}^{\text{down}} \tilde{Q} \tilde{D}^c \tilde{H}_d + \hat{y}^{\text{lept}} \tilde{L} \tilde{E}^c \tilde{H}_d  \right] + \text{H.c.},
\end{equation}
with the tildes denoting the scalar component of the chiral superfield in question. Similarly, averaging over the the gauge-kinetic F densities lead to soft-breaking masses for the gauginos, $\lambda^{(i)}(x)$,
\begin{equation}
L_{\text{gauge}} = \left[ \frac{1}{2} \sum_i \Tr W^{(i)\alpha} W_{\alpha}^{(i)}\right]_F - \frac{\xi^*}{2} \sum_i \Tr \lambda^{(i)} \lambda^{(i)} + \text{H.c.},
\end{equation}
where $W^{(i)\alpha}$ represent the standard model field-strength superfields. Refer to \cite{Kobakhidze:2008py} for further details of the derivation.

At some cut-off scale, $\Lambda$ -- also a parameter of the theory -- the soft-breaking terms for the minimal stochastic superspace model are:
\begin{subequations}\label{eq:bc}
\begin{eqnarray}
B_{\mu} = \xi^*,\label{eq:B}\\
A_0 = 2\xi^*,\label{eq:A}\\
m_{1/2} = \frac{1}{2}\lvert \xi \rvert,\label{eq:mhalf}\\
m_0^2 = 0.\label{eq:m0}
\end{eqnarray}
\end{subequations}
Here, $B_{\mu}$ is the bilinear Higgs soft term, $A_0$ is the trilinear universal coupling, $m_{1/2}$ is the universal gaugino mass and $m_0$ denotes the universal soft scalar mass. Thus, the minimal model in stochastic superspace represents a further constrained version of the so-called constrained MSSM (CMSSM). The analysis of this model can be summarized by stating the key phenomenological features at the extrema of acceptable cut-off scales:
\begin{itemize}
\item For $\Lambda = M_{\text{GUT}}$, the lightest supersymmetric particle (LSP) is the stau. Since this model conserves R-parity, the LSP is a stable particle. Thus, this model is excluded by experiment. 
\item For $\Lambda = M_{\text{Pl}}$, the LSP is the neutralino. A phenomenologically viable region of $\xi$-parameter space exists at this cut-off scale.
\end{itemize}

However, as this model of soft supersymmetry breaking from stochastic superspace is based on the MSSM, neutrino masses are inherently absent. In the following sections we explore the possibility of applying the stochastic superspace formalism to two of the standard mechanisms for generating neutrino masses in supersymmetric models: through R-parity violation and type-I seesaw. While doing so, we aim to find phenomenologically viable regions of parameter space that do not necessitate pushing the cut-off as high as the Planck scale, where quantum gravitational effects become significant. 

\subsection{Improvement of the Sparticle Spectrum Calculation}
\begin{figure}
\centering
\subfigure[$\Lambda = M_{\text{GUT}}, \xi = -500$ GeV.]{
\includegraphics[scale=0.87]{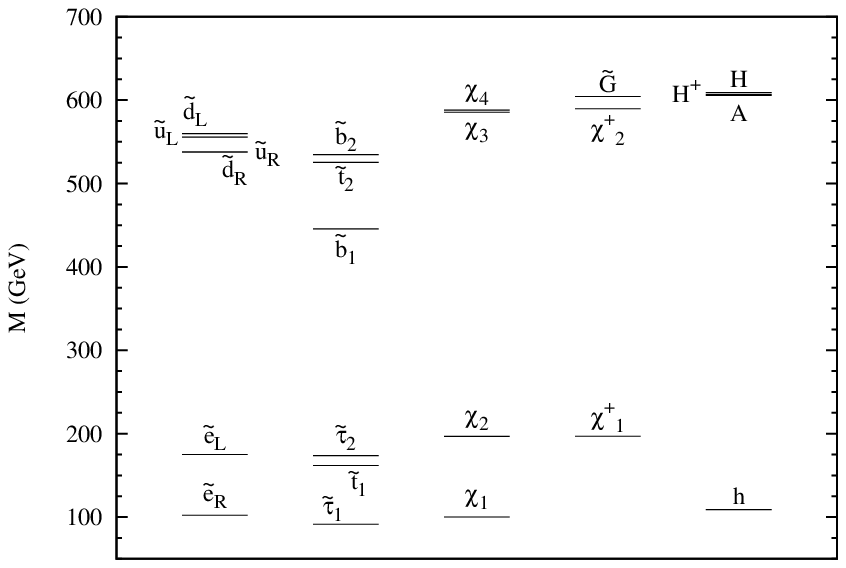}
}
\subfigure[$\Lambda = M_{\text{Pl}}, \xi = -500$ GeV.]{
\includegraphics[scale=0.87]{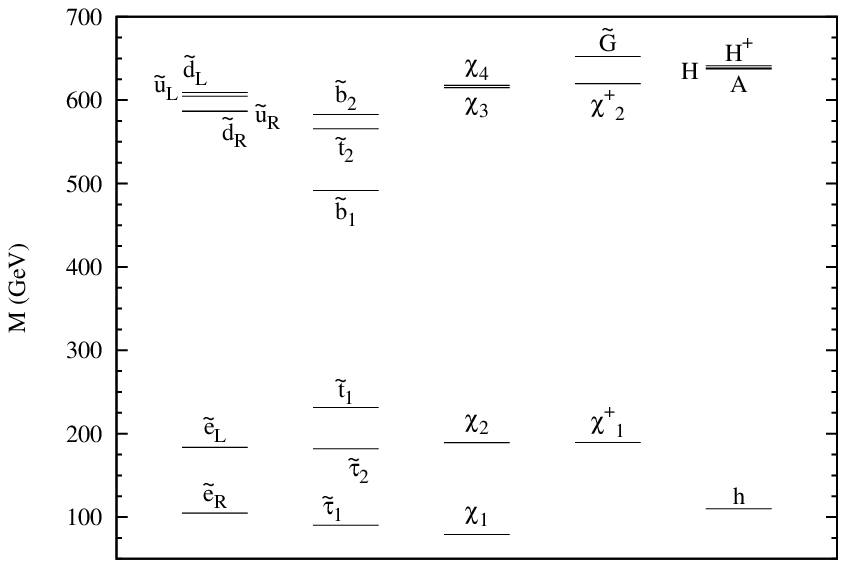}
\label{fig:compare}
}
\subfigure[$\Lambda = M_{\text{GUT}}, \xi = -750$ GeV.]{
\includegraphics[scale=0.87]{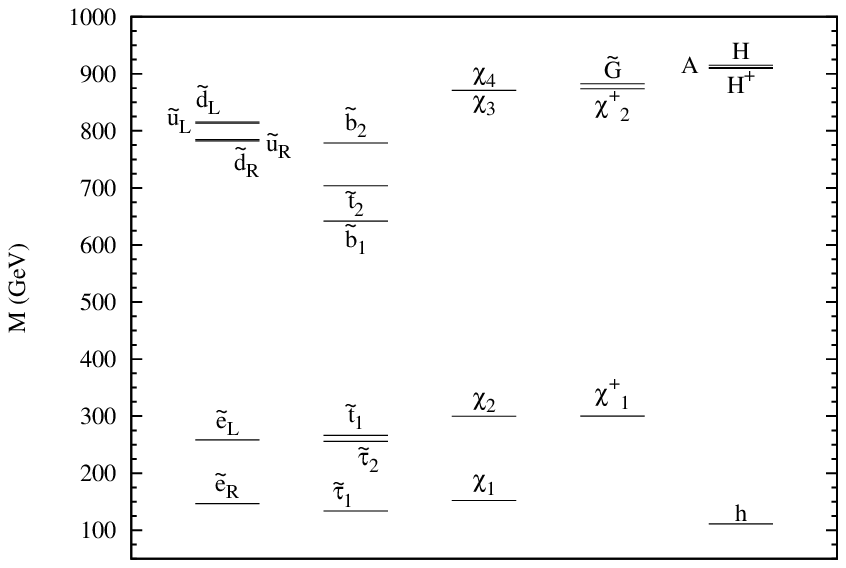}
}
\subfigure[$\Lambda = M_{\text{Pl}}, \xi = -750$ GeV.]{
\includegraphics[scale=0.87]{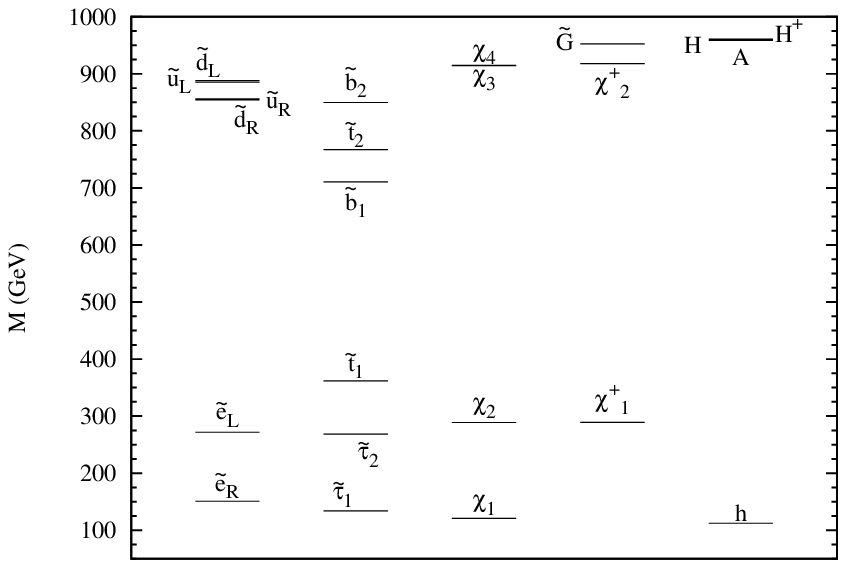}
}
\subfigure[$\Lambda = M_{\text{GUT}}, \xi = -1000$ GeV.]{
\includegraphics[scale=0.87]{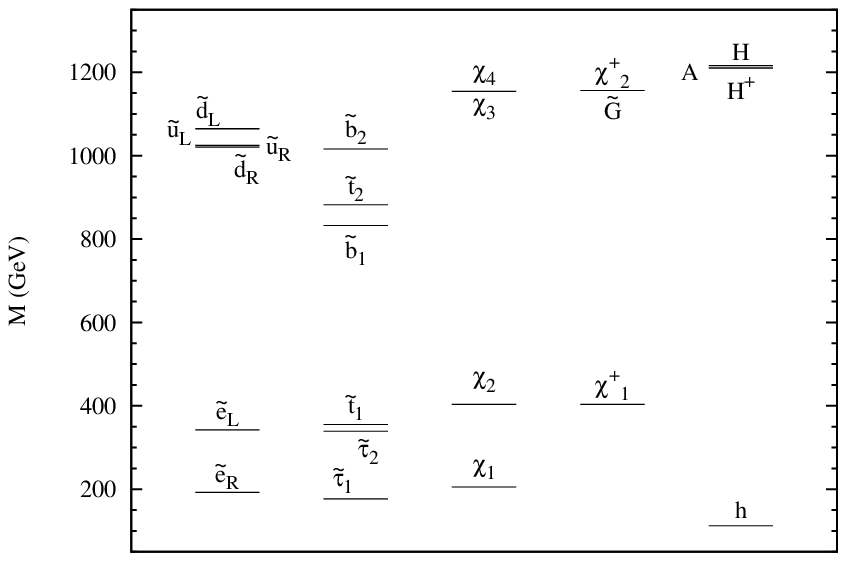}
}
\subfigure[$\Lambda = M_{\text{Pl}}, \xi = -1000$ GeV.]{
\includegraphics[scale=0.87]{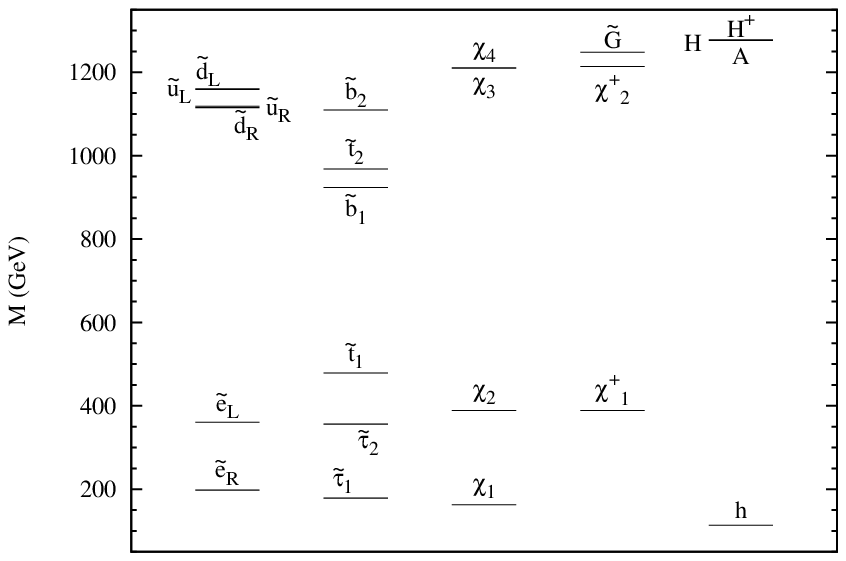}
}
\caption{Sparticle spectra computed using {\ttfamily SOFTSUSY3.0}, for various values of the parameters $\Lambda$ and $\xi$.}
\label{fig:spectra}
\end{figure}

In the original analysis of the stochastic superspace mechanism for soft supersymmetry breaking, the calculation of the superparticle spectrum was performed using an analytical solution to an approximation of the one-loop renormalization group equations \cite{Kobakhidze:2008py,Kazakov:2000us}. We have since improved upon the accuracy of these results by performing the calculations using the sparticle spectrum calculator software {\ttfamily SOFTSUSY3.0} \cite{Allanach:2001kg, Allanach:2009bv}, which uses an iterative numerical technique to solve the two-loop RGEs, whilst taking into account threshold effects. Figure \ref{fig:spectra} displays sparticle spectra calculated by {\ttfamily SOFTSUSY3.0} for various points in the $(\xi, \Lambda)$ parameter space, with Figure \ref{fig:compare} corresponding to the same parameter choice as the example spectrum presented in \cite{Kobakhidze:2008py}. Of note is a global decrease in the predicted sparticle masses, but this does not affect the generic features of the minimal stochastic superspace outlined earlier. For example, where the neutralino is LSP, there is still a large region of parameter space for which the lightest stau is no heavier than 10 GeV more than the neutralino mass. This is the stau coannihilation regime, which ensures that processes in the early universe result in the correct cosmological dark matter abundance. Additionally, the character of the LSP swaps between stau and neutralino when $\Lambda \sim 6 \times 10^{17}$ GeV, depending slightly on $\xi$.\footnote{In the literature the region of parameters with $m_0=0$ is usually considered to be excluded experimentally, due to the fact that relatively light stau is the LSP. This is simply because the soft-breaking parameters are customarily defined at $\Lambda=M_{\rm GUT}$. Ellis \emph{et al.} have also recently noticed in \cite{Ellis:2010ip} that this region of parameters is acceptable with higher cut-off scales in the CMSSM, consistent with our earlier results in \cite{Kobakhidze:2008py}.} This is also consistent with our earlier results.

As has been noted earlier, the minimal model with stochastic supersymmetry is a particular case of the CMSSM. Therefore, besides sharing generic phenomenological features with the CMSSM, our model gives more specific predictions. Namely, $\tan\beta$ is no longer a free parameter, but predicted to be $\tan\beta\approx 4\div 5$ (depending on $\xi$ and $\Lambda$). Despite such a low $\tan\beta$, the mass of the lightest CP-even Higgs boson is predicted to be $m_h\approx 112\div 115$ GeV due to the enhanced loop contribution stemming from the large mixing parameter in the stop sector.\footnote{ Taking into account typical $3\div5$ GeV theoretical error in determining $m_h$ \cite{Allanach:2004rh}, our prediction is consistent with the LEP II lower bound on SM-like Higgs boson, $m_h^{\rm LEP}> 114.4$ GeV}

Another characteristic feature of the stochastic superspace sparticle spectra is a lighter-than-usual stop mass.\footnote{Light stops are welcome within the context of electroweak baryogenesis, since they trigger a strong first-order electroweak phase transition \cite{Balazs:2004bu}.} When the cut-off scale is close to $M_{\text{Pl}}$, the stop decays predominantly through the process,
\begin{equation}
\label{eq:top}
\tilde{t}_1 \rightarrow t \tilde{\chi}_1^0.
\end{equation}
If, say, the LHC were to detect a larger cross-section for top production than expected, it could suggest stochastic superspace with a high cut-off scale. While this is true for $\xi < -500$ GeV, as $\xi$ becomes more positive the stop decays primarily through the also experimentally interesting channel,
\begin{equation}
\label{eq:b}
\tilde{t}_1 \rightarrow b \tilde{\chi}_1^+.
\end{equation}
The stochastic superspace hypothesis (at least the minimal model) can be falsified if none of these signatures are observed in future experiments. However, if (\ref{eq:top}) is observed at LHC, one might be able to extract the important stop mixing angle $\theta_t$ from the measurements of forward-backward asymmetries in leptonic and hadronic decays of polarized top quark \cite{Perelstein:2008zt}. Also, simultaneous analyses of decays (\ref{eq:top}) and (\ref{eq:b}) will allow $\theta_t$ to be determined with high accuracy \cite{Rolbiecki:2009hk}. Our model predicts: $\theta_t\approx 0.3\div 0.4$.\footnote{$|\theta_t|<\pi/4$ is defined as in \cite{Allanach:2001kg}.}

Finally, we have analyzed the amount of fine-tuning in our model by computing 
\begin{equation}
\Delta={\rm max}\left(\biggl\lvert \frac{\partial \log m_Z^2}{\partial \log p_i}\biggr\rvert \right),
\end{equation}
where $p_i$ is a relevant parameter \cite{Barbieri:1987fn}. The fine-tuning measure is found in the range $\Delta\approx 200\div 900$, with the lower limit corresponding to $\xi = -500\rm{\,GeV}$ and the upper limit corresponding to $\xi = -1000\rm{\,GeV}$. This means that the fine-tuning is $2\% \div 10\%$ over the allowed $(\xi, \Lambda)$ parameter space.

\section{R-parity Violation}
\label{sec:Rpar}
In supersymmetric extensions of the standard model it is generally possible to find renormalizable, gauge-invariant interactions that violate either lepton number or baryon number conservation. This is particularly problematic when superpotential terms that violate lepton and baryon number conservation are simultaneously present, as this will lead to disastrously rapid decay of the proton. The MSSM imposes a further symmetry, called R-parity \cite{Fayet:1977yc, Farrar:1978xj}, to forbid such dangerous phenomenological effects from arising. R-parity is a discrete symmetry defined by
\begin{equation}
R_p = (-1)^{2S + 3B + L},
\end{equation}
so that standard model particles are R-parity even ($R_p = +1$) and their superpartners are R-parity odd ($R_p = -1$). When R-parity is conserved, R-parity odd particles can only be produced in pairs, and sparticle decays must produce an odd number of sparticles. The phenomenological benefits of R-parity conservation are thus two-fold: lepton number and baryon number are conserved, and stability of the lightest supersymmetric particle (LSP) is guaranteed, which often provides a natural dark matter candidate. 

From a theoretical standpoint, however, the choice of R-parity conservation is arbitrary. If we permit R-parity to be violated, the superpotential should be augmented by the terms,
\begin{equation}
\label{eq:rpvsuperp}
W_{\slashed{R}_p} = \epsilon_{mn}\left[\mu_{i}L_{i}^m H_2^n + \frac{1}{2} \lambda_{ijk}L_{i}^m L_{j}^n E^c_k + \lambda'_{ijk} L_{i}^m Q_j^n D^c_k \right]  + \frac{1}{2} \epsilon^{\alpha \beta \gamma} \lambda''_{ijk} U^c_{i \alpha} D^c_{j \beta} D^c_{k \gamma},
\end{equation}
where $\{m,n \}$ are $SU(2)_L$ indices and $\{\alpha, \beta, \gamma \}$ are $SU(3)$ gauge indices. The first three terms violate lepton number conservation, and originate from replacing $H_1$ in the R-parity conserving superpotenial with $L_i$, which has identical gauge quantum numbers. The last term violates baryon number, and is the only remaining renormalizable and gauge-invariant superpotential term allowed when considering the same field content and gauge symmetries as the MSSM. 

It is still possible to avoid inconvenient effects such as the aforementioned proton decay if a less constraining symmetry -- baryon triality -- is considered \cite{Ibanez:1991pr}. Here, the last term in Eq.\eqref{eq:rpvsuperp} is forbidden, effectively allowing only lepton number violation. One of the key phenomenological differences resulting from R-parity violation is the generation of neutrino masses and mixings. Additionally, since the LSP can decay into standard model particles, its charge or colour neutrality is irrelevant because it no longer presents as a dark matter candidate. 

\subsection{RPV and Neutrino Masses}
The presence of bilinear R-parity violating terms in the superpotential leads to an interaction between leptons and higgsinos that was otherwise absent in the MSSM. These new interactions require the neutralino mass matrix to be extended to a $7\times 7$ matrix, which includes the three neutrinos in addition to the four neutralinos \cite{Allanach:2003eb}. The structure is reminiscent of the seesaw mass matrix, and the same techniques can be used to extract the $3\times 3$ neutrino mass matrix. The tree-level neutrino mass matrix,
\begin{eqnarray}
[m_{\text{tree-level}}]^{(\mu \mu)}_{ij} \approx  \frac{\cos^2{\beta}}{M_{\text{SUSY}}} \mu_i \mu_j,
\end{eqnarray}
results in a single non-zero eigenvalue. $M_{\text{SUSY}}$ denotes the scale at which supersymmetry is broken. Solar and atmospheric data for neutrino mass-squared differences \cite{PDG:2008} indicate that at least two neutrinos should be massive. This rules out the possibility of purely bilinear RPV interactions generating neutrino masses consistent with observations. Trilinear RPV superpotential terms couple a scalar field with two fermionic fields, resulting in radiatively generated neutrino masses from slepton-lepton and squark-quark loops. For purely trilinear effects\footnote{The bilinear RPV interactions can also generate loop contributions which may dominate under certain conditions. See \cite{Rakshit:2004rj} for a summary of other possibilities.}, the loop contributions can be expressed as \cite{Rakshit:2004rj},
\begin{align}
\label{eq:loop1}
[m_{\text{loop}}]^{(\lambda \lambda)}_{ij} \approx \sum_{k,l} \frac{1}{8 \pi^2}\lambda_{ikl} \lambda_{jlk} \frac{m_{l_k} m_{l_l}}{M_{\text{SUSY}}},\\
\label{eq:loop2}
[m_{\text{loop}}]^{(\lambda' \lambda')}_{ij} \approx\sum_{k,l} \frac{3}{8 \pi^2}\lambda'_{ikl} \lambda'_{jlk} \frac{m_{d_k} m_{d_l}}{M_{\text{SUSY}}},
\end{align}
where $m_{l_k}$ and $m_{d_k}$ are the $k^{\text{th}}$ generation charged-lepton and down-quark masses, respectively. The neutrino mass matrix is then the combination of the bilinear tree-level contributions with the trilinear loop corrections,
\begin{equation}
\label{eq:massmat}
[m_{\nu}]_{ij} = [m_{\text{tree-level}}]_{ij}^{(\mu \mu)} + [m_{\text{loop}}]_{ij}^{(\lambda \lambda)} + [m_{\text{loop}}]_{ij}^{(\lambda' \lambda')},
\end{equation}
which can, in general, yield three non-zero eigenvalues. In order to find an R-parity violating model that produces neutrino masses and mixings consistent with experimental data, we choose to diagonalize the mass matrix in Eq.\eqref{eq:massmat} with the tribimaximal mixing matrix, defined by 
\begin{equation}
U = 
\begin{pmatrix}
\sqrt{\frac{2}{3}} & \frac{1}{\sqrt{3}} & 0 \\
-\frac{1}{\sqrt{6}} & \frac{1}{\sqrt{3}} & -\frac{1}{\sqrt{2}}\\
-\frac{1}{\sqrt{6}} & \frac{1}{\sqrt{3}} & \frac{1}{\sqrt{2}}
\end{pmatrix},
\end{equation}
using the phase convention of \cite{Altarelli:2009wt}. The neutrino mass matrix can then be written in terms of the mass eigenvalues,
\begin{eqnarray}
m_{\nu} &=& U \diag(m_1, m_2, m_3) U^T \notag \\
 &=& \begin{pmatrix}
 \frac{1}{3}\left(2 m_1 + m_2\right) & \frac{1}{3}\left(-m_1+m_2\right) & \frac{1}{3}\left(-m_1+m_2\right)\\
 \frac{1}{3}\left(-m_1+m_2\right) & \frac{1}{6}\left(m_1 + 2 m_2 + 3 m_3\right) & \frac{1}{6}\left(m_1 + 2 m_2 - 3 m_3\right) \\
 \frac{1}{3}\left(-m_1+m_2\right) & \frac{1}{6}\left(m_1 + 2 m_2 - 3 m_3\right) & \frac{1}{6}\left(m_1 + 2 m_2 + 3 m_3\right) 
 \end{pmatrix}.
 \label{eq:lhs}
\end{eqnarray}
Setting this equal to the expression in Eq.\eqref{eq:massmat} defines sets of R-parity violating couplings at the weak scale that give rise to experimentally consistent neutrino masses and mixings, provided an appropriate scale for the neutrino masses is chosen. 

As our goal is to establish whether or not a phenomenologically permissible realization of R-parity violation is possible in the stochastic superspace framework, we proceed to analyze a constrained sub-section of the vast parameter space offered by the RPV regime. Contingent on a positive result, we argue that a thorough scan of a more relaxed parameter space could map further regions compatible with stochastic superspace. However, a study of this magnitude is beyond the scope of this work. 

For simplicity, we set one neutrino mass (which fixes the value of the remaining two from solar and atmospheric data) and all bilinear couplings at weak scale to zero\footnote{Due to the nature of the RGEs for $\mu_i$, $\mu_i(M_z)=0$ does not imply the same condition holds at some higher cut-off scale. Refer to the review \cite{Barbier:2004ez} for a detailed explanation.}. Inspection of the loop contributions in Eq.\eqref{eq:loop1} and Eq.\eqref{eq:loop2} reveals a dependence on products of lepton or down-quark masses, respectively. Given the mass hierarchies in these sectors, we make the approximations $m_{\tau}\gg m_{\mu},m_e$ and $m_b \gg m_s, m_d$, which simplifies the loop contributions to
\begin{equation}
[m_{\text{loop}}]^{(\lambda \lambda)}_{ij} \approx \frac{1}{8 \pi^2} \frac{m_{\tau}^2}{M_{\text{SUSY}}} \lambda_{i33} \lambda_{j33}
\end{equation}
and
\begin{equation}
[m_{\text{loop}}]^{(\lambda' \lambda')}_{ij} \approx \frac{3}{8 \pi^2} \frac{m_{b}^2}{M_{\text{SUSY}}} \lambda'_{i33} \lambda'_{j33}.
\end{equation}
Effectively, the relevant trilinear R-parity violating parameter set has been reduced to\footnote{$\lambda_{333}=0$ from antisymmetry in the first two indices.}
\begin{equation}
\label{eq:pars}
\{\lambda_{133}, \lambda_{233}, \lambda'_{133}, \lambda'_{233}, \lambda'_{333} \}. 
\end{equation}
The five resulting independent linear equations can then be solved for the set of weak scale trilinear RPV couplings that achieve neutrino masses in the normal or inverted hierarchy with tribimaximal mixing. See Table \ref{fig:trilinears} for the eight possible parameter sets for the case where one neutrino mass is zero. 
\begin{table}
\centering
\begin{tabular}{|c|c|cccc|c|cccc|}
\hline
Parameter	&	\multicolumn{5}{c|}{Normal Hierarchy}								&	\multicolumn{5}{c|}{Inverted Hierarchy} \\
\hline \hline		
$\lambda_{133}$							&	$8.058 \times 10^{-5}$ & $-$ & $+$ & $-$ & $+$	&	$3.503 \times 10^{-4}$ & $-$ & $+$ & $-$ & $+$ \\
$\lambda_{233}$							&	$1.612 \times 10^{-4}$ & $-$ & $+$ & $-$ & $+$	&	$0$ &  &  &  & \\
$\lambda'_{133}$						&	$6.802 \times 10^{-6}$ & $-$ & $-$ & $+$ & $+$	&	$6.339 \times 10^{-7}$ & $-$ & $-$ & $+$ & $+$\\
$\lambda'_{233}$						&	$5.035 \times 10^{-5}$ & $+$ & $+$ & $-$ & $-$	&	$6.077 \times 10^{-5}$ & $-$ & $-$ & $+$ & $+$\\
$\lambda'_{333}$						&	$6.395 \times 10^{-5}$ & $-$ & $-$ & $+$ & $+$	&	$6.077 \times 10^{-5}$ & $-$ & $-$ & $+$ & $+$\\
\hline
\end{tabular}
\caption{R-parity violating parameter sets (at weak scale) that achieve tribimaximal mixing in the normal hierarchy ($m_1=0$, $m_2=8.75\times 10^{-12}\,\text{GeV}$, $m_3=4.90\times 10^{-11}\,\text{GeV}$) and the inverted hierarchy ($m_1=4.90\times 10^{-11}\,\text{GeV}$, $m_2=4.98\times 10^{-11}\,\text{GeV}$, $m_3=0$). The columns of `$+$' and `$-$' represent the eight possible cases, indicating the sign of the corresponding parameter whose magnitude does not change for a given choice of mass hierarchy.}
\label{fig:trilinears}
\end{table}

\subsection{Results}
Determining the phenomenology of soft supersymmetry breaking from stochastic superspace in the context of R-parity violation follows the same procedure as for the R-parity conserving case: set the renormalization group equations' boundary conditions at some scale $\Lambda$ using the conditions in Equations \eqref{eq:bc} and run down to the weak scale to obtain the sparticle spectrum. We encounter a complication in the R-parity violating case in that we require a very specific set of RPV trilinear couplings at the weak scale to obtain realistic neutrino masses and mixings. These weak scale values must be translated to a set of cut-off scale inputs. 

We find empirically, using the RPV sparticle spectrum calculator software {\ttfamily SOFTSUSY3.0} \cite{Allanach:2009bv}, that there is an approximately linear relationship between the weak and cut-off scale values of the trilinear couplings at the order of magnitude of interest for neutrino mass generation. The proportionality constant differs for charged-lepton and down-quark couplings, and is also affected by the mSUGRA cut-off scale inputs. It must therefore be recalculated at each point in the $(\xi,\Lambda)$ parameter space. We define,
\begin{eqnarray}
\lambda(M_z) & \equiv & b \lambda(\Lambda), \\
\lambda'(M_z) & \equiv & b' \lambda'(\Lambda),
\end{eqnarray}
where the constants $b$ and $b'$ are found by setting all but one of the RPV couplings to zero. We do not discriminate between individual couplings within the charged-lepton and down-quark sectors since the difference is negligibly small. It is then straightforward to find the cut-off scale input values for the trilinear R-parity violating couplings. Non-zero trilinear couplings, however, generate contributions to the bilinear RPV couplings through the running of the RGEs. The contributions from the lepton and down-quark couplings are approximately independent, so we can express the bilinear coupling at weak scale as a linear combination of the trilinear contributions,
\begin{equation}
\label{eq:tricomb}
\mu_i(M_z) = a \lambda_{i33}(\Lambda) + a' \lambda'_{i33}(\Lambda),
\end{equation}
where the constants $a$ and $a'$ are defined by the ratios,
\begin{eqnarray}
a & \equiv & \frac{\mu(M_z)}{\lambda(\Lambda)},\\
a' & \equiv & \frac{\mu(M_z)}{\lambda'(\Lambda)}.
\end{eqnarray}
This would result in non-zero bilinear RPV couplings at the weak scale, which would then affect our previous analysis of the neutrino masses, where we have neglected the contributions from these couplings by assuming they are non-existent at the weak scale. In the absence of trilinear couplings, we notice that, like the trilinear couplings, there is an approximately linear relationship between the weak and cut-off scale values such that,
\begin{equation}
\mu(M_z) \equiv d \mu(\Lambda). 
\end{equation}
It is conceivable, then, to engineer a situation where the bilinear couplings at the cut-off scale have been chosen so that the contribution from the trilinear couplings in Eq.\eqref{eq:tricomb} is exactly cancelled at the weak scale, i.e., we set
\begin{equation}
\mu_i(\Lambda) = - \frac{a \lambda_{i33}(\Lambda) + a' \lambda'_{i33}(\Lambda)}{d}.
\end{equation}

\begin{figure}
\centering
\subfigure[$\Lambda = M_{\text{GUT}}$. For $\xi \lesssim -300$ GeV, the stau is the LSP.]{
\includegraphics[scale=0.87]{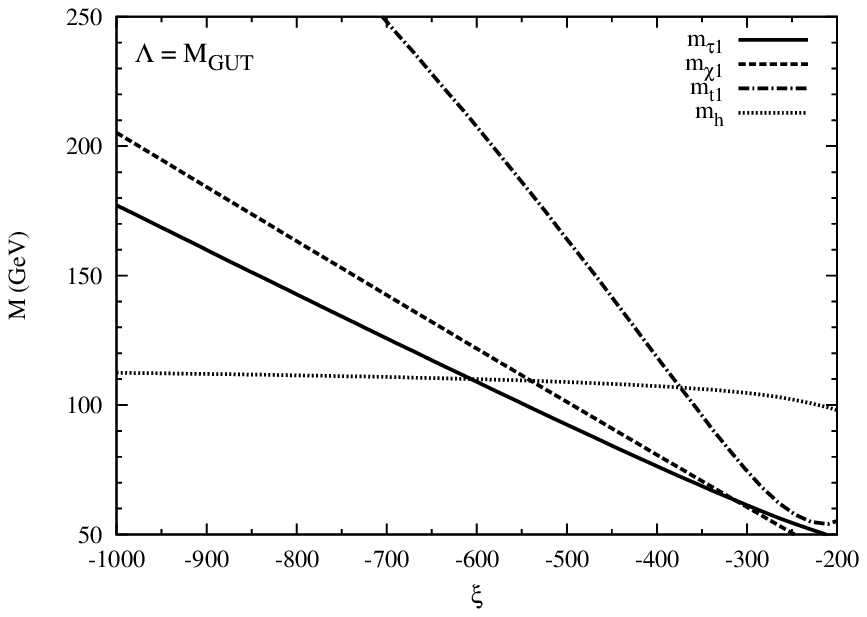}
\label{fig:gut}
}
\subfigure[$\Lambda = M_{\text{Pl}}$. The neutralino is the LSP.]{
\includegraphics[scale=0.87]{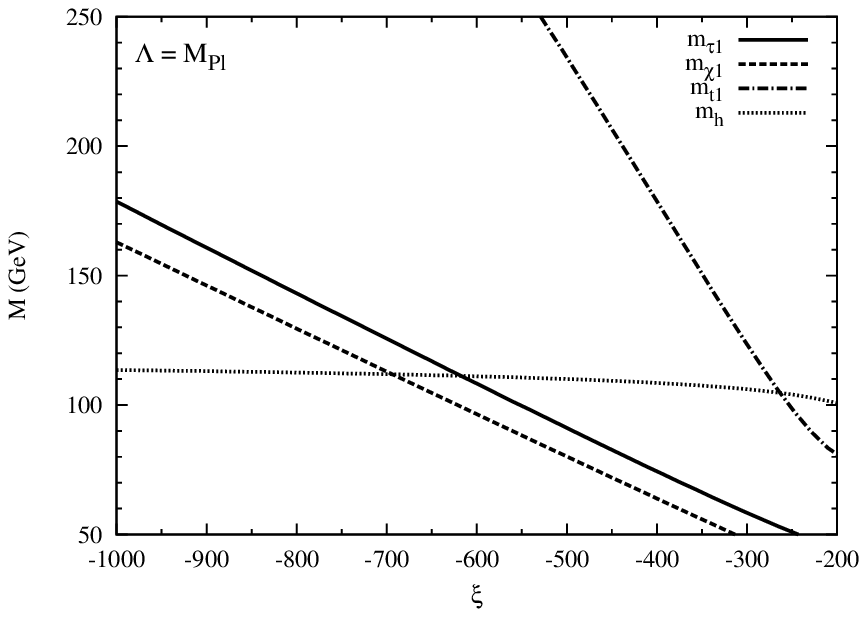}
\label{fig:planck}
}
\caption{These plots display the $\xi-$dependence of the masses of the lightest stau $(m_{\tau 1})$, lightest neutralino $(m_{\chi 1})$, lightest stop $(m_{t 1})$ and lightest CP-even Higgs for two representative choices of the cut-off scale, $\Lambda$.}
\label{fig:LSP}
\end{figure}

Using a modified version of {\ttfamily SOFTSUSY3.0.11} and the above algorithm for determining the appropriate RPV couplings at the cut-off scale, we scanned the stochastic superspace parameter space in the region $-1000 < \xi < -100$ and $M_{\text{GUT}} < \Lambda < M_{\text{Pl}}$. Because of the small magnitude of the RPV couplings, their contributions to sparticle masses through new terms in the RGEs\footnote{See the appendix of \cite{Allanach:2003eb} for a full listing of the RPV RGEs used in the {\ttfamily SOFTSUSY3.0} software.} are negligible, so the sparticle spectra for $R_p$ conserving and $R_p$ violating stochastic superspace models are not distinguishable (see Figure \ref{fig:spectra}). Like in the minimal stochastic superspace model, we find that the stau is LSP for $\Lambda = M_{\text{GUT}}$, and as the cut-off scale is pushed higher towards the Planck scale, it changes to the neutralino; see Figure \ref{fig:LSP}. 
The crucial difference is that in the R-parity violating scenario the LSP is no longer stable, so a stau LSP is not automatically excluded by experiment. It decays into Standard Model particles predominantly through the two-body processes \cite{Dreiner:2008rv},
\begin{eqnarray}
\tilde{\tau}_1^- \rightarrow \tau^- \bar{\nu}_i,\\
\tilde{\tau}_1^- \rightarrow \tau^- \nu_i,\\
\tilde{\tau}_1^- \rightarrow l_i^- \nu_{\tau}.
\end{eqnarray} 
The branching ratio for each of these processes is equal, with the 2-body decay width given by
\begin{equation}
\Gamma_{\text{two-body}} \approx \sum_i \frac{3 \lvert \lambda_{i33} \rvert^2 m_{\tilde{\tau}}}{16 \pi},
\end{equation}
which puts an upper bound on the lifetime of the stau LSP of the order $10^{-18}$s, guaranteeing its decay occurs inside the detector. Contrasting to the decay channels in Eqs.\eqref{eq:top}-\eqref{eq:b}, when R-parity is violated the stop can decay directly into Standard Model particles through the couplings $\lambda'_{i33}$. The dominating process,
\begin{equation}
\tilde{t}_1 \rightarrow b \tau,
\end{equation}
should be within experimental capability for detection at the LHC.

Flavor-changing processes are often an issue with R-parity violating models. However, due to the specific flavor structure of the R-parity violating couplings detailed in Eq.\eqref{eq:pars}, interesting flavor-changing processes are not induced at tree-level. Many are induced indirectly through the generation of relevant couplings due to the RG running. However, the couplings generated in this way are smaller than the ones in Table \ref{fig:trilinears} and thus they easily satisfy existing experimental bounds. There are also some radiative flavor-changing processes which are directly generated in our model. For example, in the model with normal neutrino mass hierarchy, the dominant contribution to the radiative decay $\mu\to e\gamma$ comes from the $\lambda$ couplings. The branching ratio is estimated as:
\begin{equation}
{\rm Br}(\mu \to e\gamma)\approx \frac{\alpha|\lambda_{133}\lambda_{233}|^2}{768\pi (G_{\rm F}m_{\tilde e_L}^2)^2}\left(2-\frac{m_{\tilde e_L}^2}{m_{\tilde e_R}^2}\right)^2\sim 10^{-21}\times\left(\frac{200~{\rm GeV}}{m_{\tilde e_L}}\right)^4.
\end{equation}
Here $\alpha\approx 1/137$ is the fine-structure constant and $G_F\approx 1.166\cdot 10^{-5}$ GeV$^{-2}$ is the Fermi constant. In the case of inverted hierarchy, $\lambda$ couplings do not contribute directly and the dominant contribution is given by $\lambda'$ couplings. We find the branching ratio to be:
\begin{equation}
{\rm Br}(\mu \to e\gamma)\approx \frac{3\alpha|\lambda'_{133}\lambda'_{233}|^2}{16\pi (G_{\rm F}m_{\tilde b_R}^2)^2}\sim  10^{-24}\times\left(\frac{600~{\rm GeV}}{m_{\tilde b_R}}\right)^4.
\end{equation}
In both cases the predicted branching ratios are many orders of magnitude smaller than, not only the current experimental upper limits, but also upper limits on the branching ratio ($\sim 10^{-14}$) which can be reached in future experiments. A similar situation pertains to other flavour-changing processes such as $\mu \to eee$ decay and $\mu \to e$ conversion in nuclei.   

We conclude that the R-parity violating extension to the minimal stochastic superspace model is phenomenologically viable, producing neutrino masses and mixings consistent with experiment whilst keeping the magnitude of the RPV couplings small enough that they do not conflict with published constraints \cite{Abada:2000xr, Barbier:2004ez}. The only outstanding issue is the removal of a dark matter candidate from the theory. This can be remedied by, for example, the introduction of axions and axinos. We do not address this here, but aim to explore the details of RPV dark matter with stochastic superspace in future work.

\section{Type-I Seesaw}
\label{sec:seesaw}
Another well-motivated way to introduce neutrino masses to a supersymmetric model is through the seesaw mechanism, of which there are three types. We base the following analysis on the simplest of the three: the type-I seesaw mechanism. For type-I seesaw, the field content of the R-parity conserving MSSM is extended by three additional gauge singlet chiral superfields, $N$, for the right-handed neutrinos which we assume to be degenerate. The seesaw superpotential,
\begin{equation}
W_{\text{seesaw}} = \mu H_u H_d + \hat{y}^{\text{up}} Q U^c H_u + \hat{y}^{\text{down}} Q D^c H_d + \hat{y}^{\text{lept}} L E^c H_d +  \hat{y}^{\text{neut}} L N^c H_u + \frac{1}{2} M_R N^c N^c,
\end{equation}
includes a neutrino Yukawa coupling and Majorana mass term that were not present in the MSSM. A neutrino Dirac mass matrix arises from the new Yukawa term, giving 
\begin{equation}
m_D = \hat{y}^{\text{neut}} \langle H_u \rangle,
\end{equation}
where $\langle H_u \rangle = v \sin \beta$ is the vacuum expectation value for the Higgs field with positive hypercharge, $v = 174$ GeV and we make the dominant third family approximation with respect to the Yukawa couplings. The light neutrino mass matrix,
\begin{eqnarray}
m_{\nu} &=& -m_D^T M_R^{-1} m_D,
\end{eqnarray}
follows from evaluating the eigenvalues of the neutrino mass matrix under the assumption that the Majorana mass scale is much greater than the Dirac mass. Adopting the notation of \cite{Kazakov:2000us}, we define the largest coupling
\begin{equation}
Y_N \equiv \frac{\left(\hat{y}^{\text{neut}}\right)^2}{16 \pi^2},
\end{equation}
which becomes a free parameter of the theory. The magnitude of $Y_N$ sets the seesaw scale,
\begin{equation}
M_R \sim Y_N \times 10^{17} \text{GeV},
\end{equation}
where we have assumed the light neutrinos' masses to be $\mathcal{O}(10^{-11}\text{GeV})$. For simplicity, in the following section we strive to generate light neutrinos with mass of this scale, without taking into account the experimentally observed mixing and mass difference data.

\subsection{Results}
Introducing the right-handed neutrino superfields affects the RGEs above the seesaw scale; see Appendix \ref{sec:appendix} for a listing of the equations that differ from the MSSM. To find the sparticle spectrum for our model with soft breaking from stochastic superspace, we set the cut-off scale boundary conditions as per Equations \eqref{eq:bc} and run the equations down to the seesaw scale, $M_R$ using the modified RGEs. Below the seesaw scale, the right-handed neutrinos decouple and the soft parameters run as in the MSSM. The neutrino Yukawa coupling remains constant in this region. 

\begin{figure}
\centering
\subfigure[$\Lambda = M_{\text{GUT}}$.]{
\includegraphics[scale=0.87]{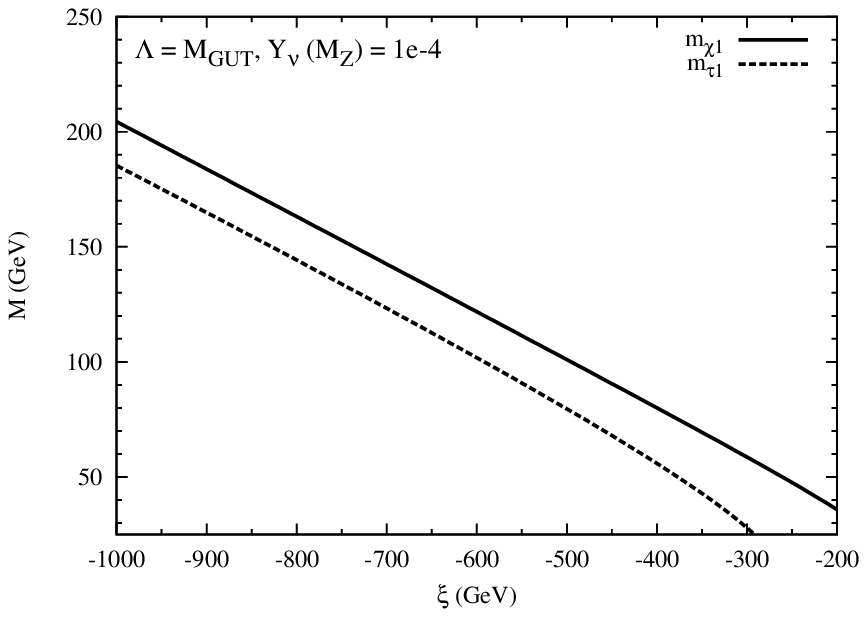}
}
\subfigure[$\Lambda = M_{\text{Pl}}$.]{
\includegraphics[scale=0.87]{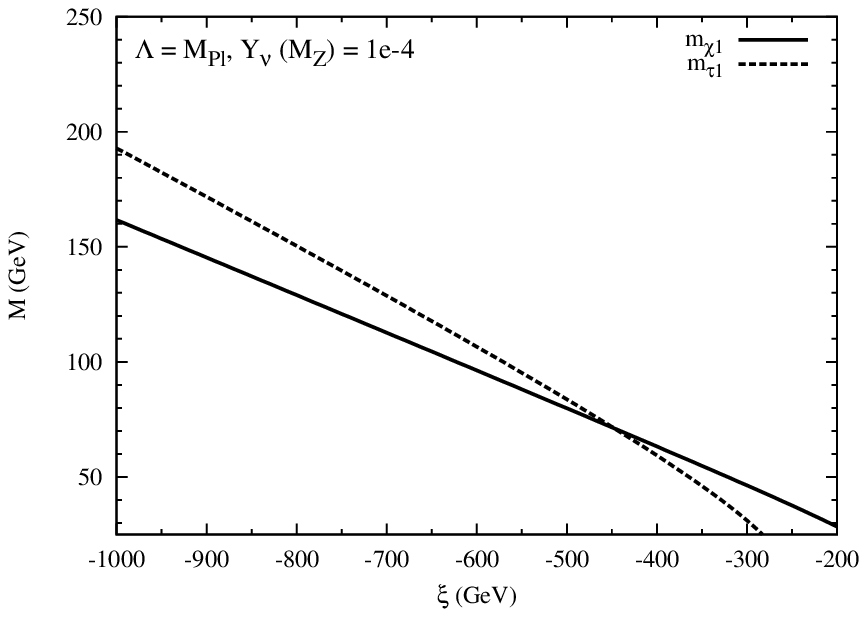}
}
\caption{LSP and NLSP masses as a function of $\xi$ for type-1 seesaw mechanism in stochastic superspace. These figures demonstrate the increased mass gap between the lightest stau and neutralino compared to the minimal stochastic superspace model.}
\label{fig:LSPseesaw}
\end{figure}

We find an upper limit on the magnitude of the neutrino Yukawa coupling at weak scale of
\begin{equation}
\label{eq:yncon}
Y_N < 0.05 Y_t.
\end{equation}
For larger values of $Y_N$, the model is ruled out for all values of the the cut-off scale, $M_{\text{GUT}} < \Lambda < M_{\text{Planck}}$, due to a stau LSP over this parameter range. The large neutrino Yukawa couplings drive the left-hand lepton masses down, which in turn lowers the mass of the lightest stau through mixing. If the neutrino Yukawa coupling is smaller than the constraint in Eq. \eqref{eq:yncon}, then for a region of $(\xi, \Lambda)$ parameter space the results of the minimal stochastic superspace model can be recovered, of course now with non-zero neutrino masses due to the seesaw mechanism. Figure \ref{fig:LSPseesaw} depicts the mass dependence of the two LSP candidates -- the lightest stau and neutralino -- on the stochasticity parameter $\xi$ at the extremes of the cutoff parameter range. For low cutoffs, the stau is the LSP by a wide margin, however as the cutoff scale is increased, some phenomenologically viable regions exist for large negative values of $\xi$. Since the slepton mass is typically lower in the seesaw model, the acceptable region of parameter space is significantly reduced compared to the minimal stochastic superspace model. 

\section{Conclusion}
\label{sec:conclusion}
The extension of minimal model with stochastic supersymmetry to incorporate neutrino masses consistent with experimental observations can successfully be achieved through R-parity violation. We found a specific set of trilinear R-parity violating couplings that lead to neutrino masses which satisfy the tribimaximal mixing scheme. The characteristic feature of the resulting sparticle spectra is lighter than usual stop masses, which could lead to interesting decay signatures at the Large Hadron Collider. However, by choosing this method to radiatively give mass to the neutrinos, a natural dark matter candidate is sacrificed. This is not necessarily problematic, as the introduction of axions and axinos can solve this issue along with the strong CP problem, which was not addressed here. The type-I seesaw mechanism is an alternative approach for generating neutrino masses. We have found that it is compatible with stochastic supersymmetry for scenarios which have $Y_N < 0.05 Y_t$ and a generally large cut-off scale. We hope to investigate the viability of type-II and type-III seesaw mechanisms for generating neutrino masses with stochastic superspace in future work.

\acknowledgments
We would like to thank Elisabetta Barberio, Anna Phan and Nicholas Setzer for useful discussions. This work was partially supported by the Australian Research Council. NP was supported by the Commonwealth of Australia.

\appendix
\section{Type-I Seesaw Renormalization Group Equations}\label{sec:appendix}
We adopt the notational conventions of \cite{Kazakov:2000us}. The RGEs are the same as for the MSSM with the following exceptions above the seesaw scale:

\begin{eqnarray}
\frac{dY_N}{dt} &=& -Y_N\left[3 a_2 + \frac{3}{5} a_1 - 3 Y_U - 4 Y_N\right], \\
\frac{dA_N}{dt} &=& 3 a_2 M_2 + \frac{3}{5} a_1 M_1 + 3 Y_U A_U + 4 Y_N A_N, \\
\frac{dY_U}{dt} &=& -Y_U\left[\frac{16}{3}a_3 + 3 a_2 + \frac{13}{15} a_1 - 6 Y_U - Y_D - Y_N\right], \\
\frac{dA_U}{dt} &=& \frac{16}{3}a_3 M_3 + 3 a_2 M_2 + \frac{13}{15} a_1 M_1 + 6 Y_U A_U + Y_D A_D + Y_N A_N, \\
\frac{dB}{dt} &=& 3 a_2 M_2 + \frac{3}{5} a_1 M_1 + 3 Y_U A_U + 3 Y_D A_D + Y_L A_L + Y_N A_N, \\
\frac{d\tilde{m}_L^2}{dt} &=& - \bigg[ 3 a_2 M_2^2 + \frac{3}{5} a_1 M_1^2 - Y_L \left(\tilde{m}_L^2 + \tilde{m}_E^2 + m_{H_1}^2 + A_L^2 \right)\notag \\
& & - Y_N \left(\tilde{m}_L^2 + \tilde{m}_N^2 + m_{H_2}^2 + A_N^2 \right) \bigg], \label{eq:LHlepton}\\
\frac{d\tilde{m}_N^2}{dt} &=& 2 Y_N \left(\tilde{m}_L^2 + \tilde{m}_N^2 + m_{H_2}^2 + A_N^2 \right),\\
\frac{d\mu^2}{dt} &=& - \mu^2 \left[3 a_2 + \frac{3}{5} a_1 - \left(3 Y_U + 3 Y_D + Y_L + Y_N\right) \right],\\
\frac{d m_{H_2}^2}{dt} &=& -\bigg[3 a_2 M_2^2 + \frac{3}{5} a_1 M_1^2 - 3 Y_U \left(\tilde{m}_Q^2 + \tilde{m}_U^2 + m_{H_2}^2 + A_U^2 \right)\notag \\
&& - Y_N \left(\tilde{m}_L^2 + \tilde{m}_N^2 + m_{H_2}^2 + A_N^2 \right) \bigg],
\end{eqnarray}
where $t = \ln \left(Q^2/ \Lambda^2 \right)$.


\begin{thebibliography}{99}
\bibitem{Kolda:1997wt}
  C.~F.~Kolda,
  Nucl.\ Phys.\ Proc.\ Suppl.\  {\bf 62}, 266 (1998)
  [arXiv:hep-ph/9707450].

\bibitem{Kobakhidze:2008py}
  A.~Kobakhidze, N.~Pesor and R.~R.~Volkas,
  Phys.\ Rev.\  D {\bf 79}, 075022 (2009)
  [arXiv:0809.2426 [hep-ph]].

\bibitem{Allanach:2009bv}
  B.~C.~Allanach and M.~A.~Bernhardt,
  arXiv:0903.1805 [hep-ph].
  
\bibitem{Kazakov:2000us}
  D.~I.~Kazakov,
  Phys.\ Rept.\  {\bf 344}, 309 (2001)
  [arXiv:hep-ph/0001257].
  
\bibitem{Allanach:2001kg}
  B.~C.~Allanach,
  Comput.\ Phys.\ Commun.\  {\bf 143} (2002) 305
  [arXiv:hep-ph/0104145].
  
\bibitem{Ellis:2010ip}
  J.~Ellis, A.~Mustafayev and K.~A.~Olive,
  arXiv:1003.3677 [hep-ph].
 
\bibitem{Allanach:2004rh}
  B.~C.~Allanach, A.~Djouadi, J.~L.~Kneur, W.~Porod and P.~Slavich,
  JHEP {\bf 0409} (2004) 044
  [arXiv:hep-ph/0406166].

\bibitem{Balazs:2004bu}
  C.~Balazs, M.~S.~Carena and C.~E.~M.~Wagner,
  Phys.\ Rev.\  D {\bf 70} (2004) 015007
  [arXiv:hep-ph/0403224].
 
\bibitem{Perelstein:2008zt}
  M.~Perelstein and A.~Weiler,
  JHEP {\bf 0903} (2009) 141
  [arXiv:0811.1024 [hep-ph]].

\bibitem{Rolbiecki:2009hk}
  K.~Rolbiecki, J.~Tattersall and G.~Moortgat-Pick,
  arXiv:0909.3196 [hep-ph].

\bibitem{Barbieri:1987fn}
  R.~Barbieri and G.~F.~Giudice,
  Nucl.\ Phys.\ B {\bf 306} (1988) 63.

\bibitem{Fayet:1977yc}
  P.~Fayet,
  Phys.\ Lett.\  B {\bf 69}, 489 (1977).
 
\bibitem{Farrar:1978xj}
  G.~R.~Farrar and P.~Fayet,
  Phys.\ Lett.\  B {\bf 76}, 575 (1978).
  
\bibitem{Ibanez:1991pr}
  L.~E.~Ibanez and G.~G.~Ross,
  Nucl.\ Phys.\  B {\bf 368}, 3 (1992).
 
\bibitem{Allanach:2003eb}
  B.~C.~Allanach, A.~Dedes and H.~K.~Dreiner,
  Phys.\ Rev.\  D {\bf 69}, 115002 (2004)
  [Erratum-ibid.\  D {\bf 72}, 079902 (2005)]
  [arXiv:hep-ph/0309196].

\bibitem{PDG:2008}  
	C.~Amsler~\emph{et~al}. (Particle Data Group), Physics Letters {\bf B667}, 1 (2008) 
  
\bibitem{Rakshit:2004rj}
  S.~Rakshit,
  Mod.\ Phys.\ Lett.\  A {\bf 19}, 2239 (2004)
  [arXiv:hep-ph/0406168].
  
\bibitem{Altarelli:2009wt}
  G.~Altarelli,
  arXiv:0905.3265 [hep-ph].

\bibitem{Barbier:2004ez}
  R.~Barbier {\it et al.},
  Phys.\ Rept.\  {\bf 420}, 1 (2005)
  [arXiv:hep-ph/0406039].

\bibitem{Dreiner:2008rv}
  H.~K.~Dreiner, S.~Grab and M.~K.~Trenkel,
  Phys.\ Rev.\  D {\bf 79}, 016002 (2009)
  [Erratum-ibid.\  {\bf 79}, 019902 (2009)]
  [arXiv:0808.3079 [hep-ph]].

\bibitem{Abada:2000xr}
  A.~Abada and M.~Losada,
  Phys.\ Lett.\  B {\bf 492}, 310 (2000)
  [arXiv:hep-ph/0007041].

\end{thebibliography}
\end{document}